\documentclass[twocolumn,showpacs,preprintnumbers,prl]{revtex4}
\usepackage{amssymb}
\usepackage[dvips]{graphicx}
\begin{document}
\title{Molecular Dynamics Study of a Thermal Expansion Coefficient:\\ Ti Bulk with an
Elastic Minimum Image Method}
\author{Yakup Hundur$^1$\footnote{To whom
correspondence should be addressed. E-mail: hundur@itu.edu.tr.},
Rainer Hippler$^2$ and Ziya B. G\"uven\c{c}$^3$}
\affiliation{$^1$Department of Physics, Istanbul Technical
University, Ayazaga, Istanbul, TR-34469 Turkey}
\affiliation{$^2$Institut f\"{u}r Physik,
Ernst-Moritz-Arndt-Universit\"{a}t Greifswald, Domstra{\ss}e 10a,
D-17487 Greifswald, Germany} \affiliation{$^3$Electronic and
Communication Engineering, \c{C}ankaya University, Balgat, Ankara,
TR-06530 Turkey}
\begin{abstract}
Linear thermal expansion coefficient (TEC) of Ti bulk is
investigated by means of molecular dynamics simulation. Elastic
Minimum Image Convention of periodic boundary conditions is
introduced to allow the bulk to adjust its size according to the
new fixed temperature. The TEC and the specific heat of Ti are
compared to the available theoretical and experimental data.
\end{abstract}

 \maketitle


\section{Introduction} Titanium is not present in pure form in the
nature, and its composites have different thermal expansion
coefficients (TEC) \cite{Hun1}. Determining the TEC values may
help to guess the composition. For expansion simulations, various
Molecular Dynamics (MD) methods allowing for a volume change exist
\cite{Hun2,Hun3}. Expansion is defined through a change of the
bond length which may be determined with several methods, like the
average length of the bulk diagonals \cite{Hun4}. For a more
statistically dependable result, the mean distances between the
atoms in the bulk are calculated in this work. Expansion of the
sample requires elasticity of the periodic boundary conditions
(PBC), but the PBC and its minimum image convention use fixed box
lengths in the simulation \cite{Hun2}. Therefore, elastic minimum
image convention (EMIC) of the PBC is introduced. It resembles
Berendsen's barostat method. However, Berendsen's method allows
only isotropic changes in the volume of the simulation box
\cite{Hun2}, while the EMIC method allows change of the shape as
well as its size. For the present purpose correctness of the
elastic constants and phonon frequencies are important. The recent
many-body potentials such as Finnis-Sinclair, Gupta, and Glue
potentials and embedded atom method cannot reproduce all elastic
constants correctly \cite{Hun5}, while Lennard-Jones potential
(LJ) \cite{Hun6} give good results \cite{Hun7} for Ti. Hence the
LJ potential is employed into our temperature scaled isoenergetic
and isobaric MD codes \cite{Hun4}. Relaxation runs of the bulk Ti
have been performed by using this code. The Verlet integration
algorithm \cite{Hun8} is used, since it allows for time-reversible
actions \cite{Hun9}.
\section{Elastic Minimum Image
Convention (EMIC)} The EMIC method is used to adjust the
simulation box lengths to the temperature changes during a run
\cite{Hun4}. The entire simulation time $t_{tot}$ is divided into
$nb$ number of time-blocks $Tb$, each consisting of $m$ time-steps
$ts$, with $\delta t$ time step size ($tss$), which can be
expressed as $t_{tot} = nb \times Tb$ and $Tb = m \times \delta
t$.

The mean bond length $MBL$ is calculated from
\begin{equation}
MBL = \frac{1}{C(N,2)} \sum_{i>j}^{N}r_{ij}
\label{EqHun3} \end{equation}
where $N$, $C$ and $r_{ij}$ are the total number of atoms, combination of $N$
by 2 and pair distance in the simulation cell, respectively. The
changes in $MBL$ are reflected directly to the box length.
Although, temperature scaling is applied after each $Tb$, EMIC is only
applied after some number of $Tb$, let's call $nscb$. Therefore, EMIC
is applied at every $ecb$ time interval, i.e., $ecb =  nscb \times Tb$,
where $nscb$ = [1--100]. The system temperature fluctuates freely
during each $Tb$, while
the $MBL$ differences accumulate and the system tries to reach an
equilibrium during each $ecb$ \cite{Hun4}. In this way, the system
also fulfils ergodicity. The difference in $MBL$ during the $k^{th}$ and
$(k-1)^{th}$ $ecb$ is
\begin{eqnarray}
&\Delta MBL_{k,k-1} = \left( \frac{1}{nscb}\sum_{i=1}^{nscb} \overline{MBL_i} \right )_k \cr
& - \left( \frac{1}{nscb}\sum_{i=1}^{nscb} \overline{MBL_i} \right )_{k-1} \cr
& = \langle \overline{MBL} \rangle_{ecb(k)}  - \langle \overline{MBL} \rangle_{ecb(k-1)}
\label{EqHun4}              \end{eqnarray}
where $\overline{MBL_i}$ is the average during the $i^{th}$ $Tb$, and the terms on the
right hand side are the $MBL$ averages at the $k^{th}$ and $(k-1)^{th}$ $ecb$.
An effective $MBL$ average at the $k^{th}$ $ecb$ is obtained by
multiplication of $\Delta MBL_{k,k-1}$  with an adjustable parameter $p$ as
\begin{equation}
\langle \overline{MBL} \rangle^{eff}_{ecb(k)}
= p \, \Delta MBL_{k,k-1} + \langle \overline{MBL} \rangle^{eff}_{ecb(k-1)}
 \label{EqHun5}              \end{equation}
where $0 < p \le 1$. Note that Eq. \ref{EqHun5} becomes Eq. \ref{EqHun4} if $p=1$.
Finally, the
box length of the EMIC in the $x$-direction from $k^{th}$ to $(k+1)^{th}$
$ecb$ is
\begin{equation}
BoxLx_{ecb(k+1)}=BoxLx_{ecb(k)} \, \frac{\langle \overline{MBL}
\rangle^{eff}_{ecb(k)}}{\langle \overline{MBL}
\rangle^{eff}_{ecb(k-1)}} \, . \label{EqHun6}
\end{equation} Similar equations hold for the other directions.
Since the radial $MBL$ averages are reflected to each of the
directions as in Eq. \ref{EqHun5}, the initial shape of the bulk
is forced remain to unchanged. However, the real lattice may not
expand with the same rate along each of the Cartesian directions
\cite{Hun10, Hun11}. Therefore Eq. \ref{EqHun5} is reformulated
for each coordinate separately as
\begin{equation}
\langle \overline{MBLx} \rangle^{eff}_{ecb(k)}
= p \, \Delta MBLx_{k,k-1} + \langle \overline{MBLx} \rangle^{eff}_{ecb(k-1)}
 \label{EqHun7}              \end{equation}
Similarly, Eq. \ref{EqHun6} is also reformulated for each direction separately
as
\begin{equation}
BoxLx_{ecb(k+1)}=BoxLx_{ecb(k)} \, \frac{\langle \overline{MBLx}
\rangle^{eff}_{ecb(k)}}{\langle \overline{MBLx}
\rangle^{eff}_{ecb(k-1)}}    \label{EqHun8}
\end{equation}  Values of $m$ in a $Tb$ should be chosen as large as possible for the
numerical convergence of the physical quantities, in order to reduce
non-physical effects \cite{Hun2}. Hence, several test runs are
required to determine the parameters $m$, $nscb$, and $p$.

\section{Simulation} The conventional  $\alpha$--Ti has hcp structure.
The characteristics of a hcp cell are determined by
proportionality of simple cell sizes $c/a$. This ratio and the
binding energy parameter $\varepsilon_0$ are taken as 1.62073 and
0.20154 eV when producing the LJ potential for Ti \cite{Hun7},
while the recent experimental $c/a$ value is 1.601 \cite{Hun12}.
The unit cell sizes are $a= 2.94055$ \AA~ and $c= 4.76583$ \AA~
for LJ potential compared to $a= 2.953$ \AA~ and $c= 4.729$ \AA~
from experiment. The bulk sample in the simulation is prepared
using the former set of constants. A safe cut-off radius for the
interactions is chosen as $2.8 \, \sigma$ ($\sigma=2^{1/6} r_0$)
\cite{Hun2}, where the bond length $r_0$ is 2.96292 \AA~
\cite{Hun7}. Since the box length should be equal or greater than
twice the cut-off radius \cite{Hun2}, the smallest possible sample
having 389 Ti atoms with a lattice size of $6a \times 6b \times
4c$, i.e., $17.718 \times  17.718 \times 18.916$ \AA$^3$ is
constructed.

Simulations have been carried out for the temperature
range 100--400 K, in steps of 100 K. Quantum
effects on the specific heat become important at low temperatures
\cite{Hun13}; as these effects are not included in the binary
potentials \cite{Hun14}, the lower temperature was chosen as 100 K.
The higher temperature limit is set near
the Debye's temperature, since the binary potentials require
modifications for the higher temperatures \cite{Hun3}. The final
velocities and coordinates of the previous run are used as the
initial conditions for the next run, i.e., those of the 200 K are
used as the initial values for the 300 K run, and so on. At each
temperature two different simulations were performed one after the
other. Firstly, the bulk is run under constant pressure by using
the EMIC boundary conditions to adjust the bulk sides. After that,
the code is run by using PBC under this constant volume.

The time averaged temperature over each $Tb$ is calculated from
\begin{equation}  T(K) = \frac{2 \langle E_{kin} \rangle}{(3N-6)
k_B}
 \label{EqHun9}              \end{equation} where
$\langle E_{kin} \rangle$ is the time averaged kinetic energy of
the atoms, $N$ is the number of atoms in the sample, and $k_B$ is
the Boltzmann constant. Hence, the simulation temperature is the
average over all $T(K)$ values.

A number of runs have been made to find the right
parameters. The $tss$ and $Tb$ are varied between 0.2--2 fs (1fs =
10$^{-15}$ sec) and 10--100 $tss$, respectively. The total number of time
steps lie within the range of 163,000--2,002,000 and the resulting
total run times are varied in between 18,700 fs and 402,000 fs.
The coefficient $p$ of the effective $MBL$ differences
is used as 0.5--1 in order to let the system change smoothly. Otherwise, if
$p$ is taken equal to 1, a sudden change in the periodic box size may lead
to excess energy at some particles.

Firstly, average bond
differences $\Delta MBL_{k,k-1}$ are distributed proportionally,
according to $c/a$ ratio,
to each of the Cartesian directions of the EMIC box by Eqs. \ref{EqHun5}
and \ref{EqHun6}. Therefore the shape of the bulk is enforced to remain
globally unchanged. However, this brutal sharing brought the
problem of exceeding force accumulation, and the problem of
stability at 300 K and 400 K. Using small time step sizes has
solved the problem of 300 K, but not at the higher
temperatures. Then, $\Delta MBL_{k,k-1}$  is shared to each Cartesian coordinate
separately, and the bulk is reconstructed in each direction
independently (Eqs. \ref{EqHun7} and \ref{EqHun8}). This causes accidental image
conventions of the coordinates after some number of time
intervals. In order to avoid these and to reach numerical
convergence of the physical quantities, some number of runs have
been done to determine the run parameters.

Together with the
temperature of the bulk, coordinates of the atoms along each of
the Cartesian directions, kinetic, potential and total energies of
the atoms are calculated. Fluctuations on all these quantities,
specific heat and thermal expansion coefficients are obtained.

The linear thermal expansion coefficient (TEC) is defined as
\begin{equation}  \alpha = \frac{1}{L_0} \frac{\Delta L}{\Delta T}
\label{EqHun10}              \end{equation}  where $L_0$, $\Delta
L$, and $\Delta T$ are the initial length, change of length and
change of temperature, respectively. In practice, $L$ is
calculated with various methods throughout the entire run. In this
work, $MBL$ and $MBL_0$ are used in place of $L$ and $L_0$,
respectively.

Heat capacity can be observed by doing a series of simulations at
different temperatures with corresponding constant volumes. It can
be calculated in each run separately by using temperature or
energy fluctuations. Various authors have offered similar forms
for the calculation of heat capacity \cite{Hun2, Hun9}. It is
defined through kinetic energy fluctuations for the microcanonical
ensembles by Lebowitz et al. \cite{Hun15} as
\begin{equation} C_V = \frac{3}{2} Nk_B \left(1- \frac{2}{3}
\frac{\sigma^2(KE)}{N k_B T^2} \right)^{-1}  \label{EqHun11}
\end{equation}      where $\sigma^2(KE)$  is mean square
deviation of the kinetic energy throughout the entire run.

\section{Results and Discussions} \subsection{Shape of the Bulk}
Small changes on the coordinates of the bulk atoms are observed.
Typical bond length changes are about 0.01--0.04 \AA~
(0.34--1.35\% of $T= 0$ K bond length). The centre of mass is
unchanged during the simulations. Between 100--300 K the shape of
the bulk remains more or less unchanged. However, at 400 K the
resulting bulk has a shape anomaly like in the experiments where
the expansion of the bulk varies in different directions
\cite{Hun10, Hun11}; the difference of the $c/a$ value after the
400 K simulation is roughly 0.4\% in comparison to its $T=0$ K
value.
\begin{figure}[h] 
\centering
\includegraphics[width=0.8\linewidth,angle=-0,clip]{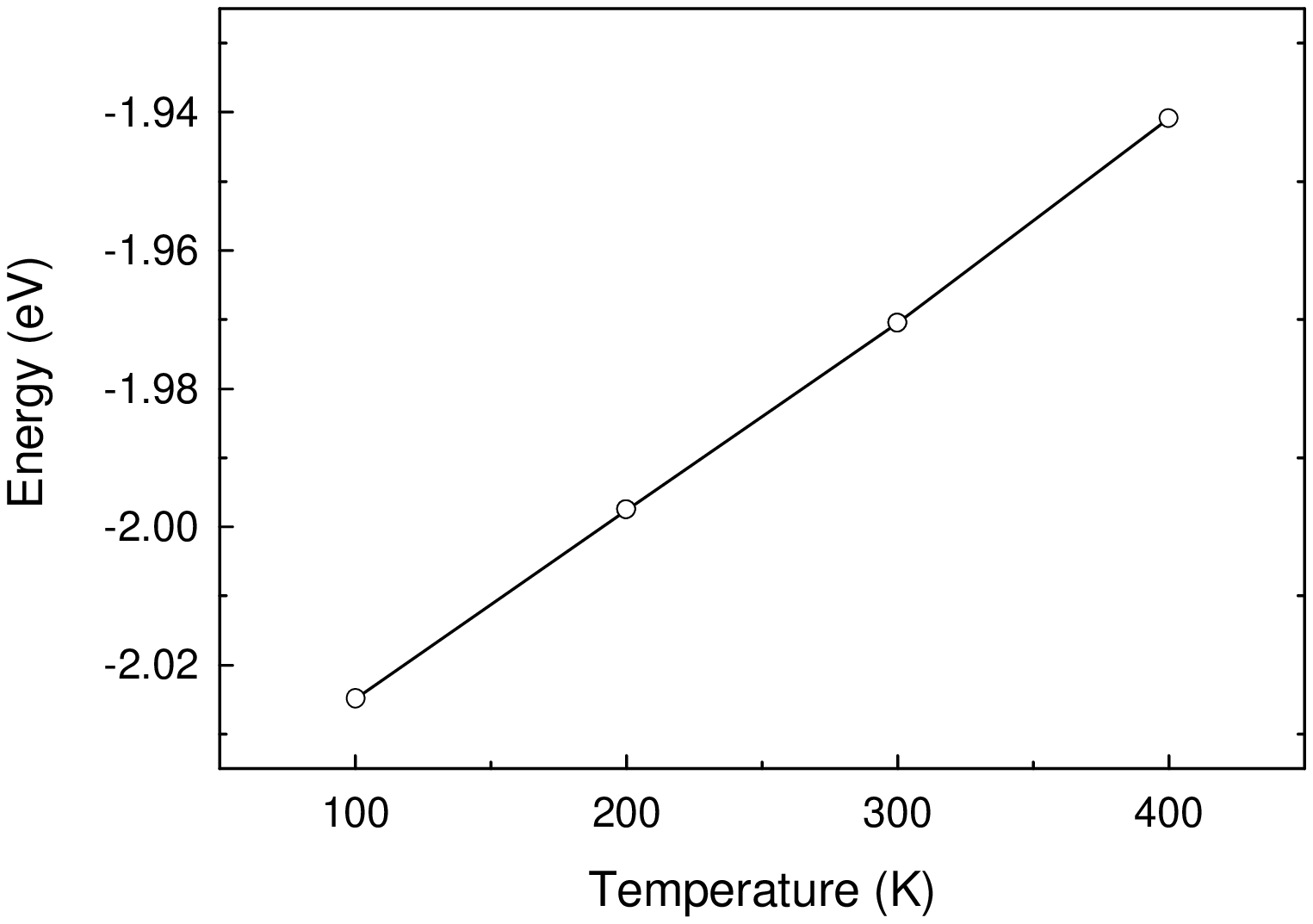}
\caption{Caloric curve of the Ti--389 bulk obtained by the
simulation.}
 \label{Hunfig1}
\end{figure}

\subsection{Specific Heat Capacity} 
\begin{figure}[h] 
\centering
\includegraphics[width=0.8\linewidth,angle=0,clip]{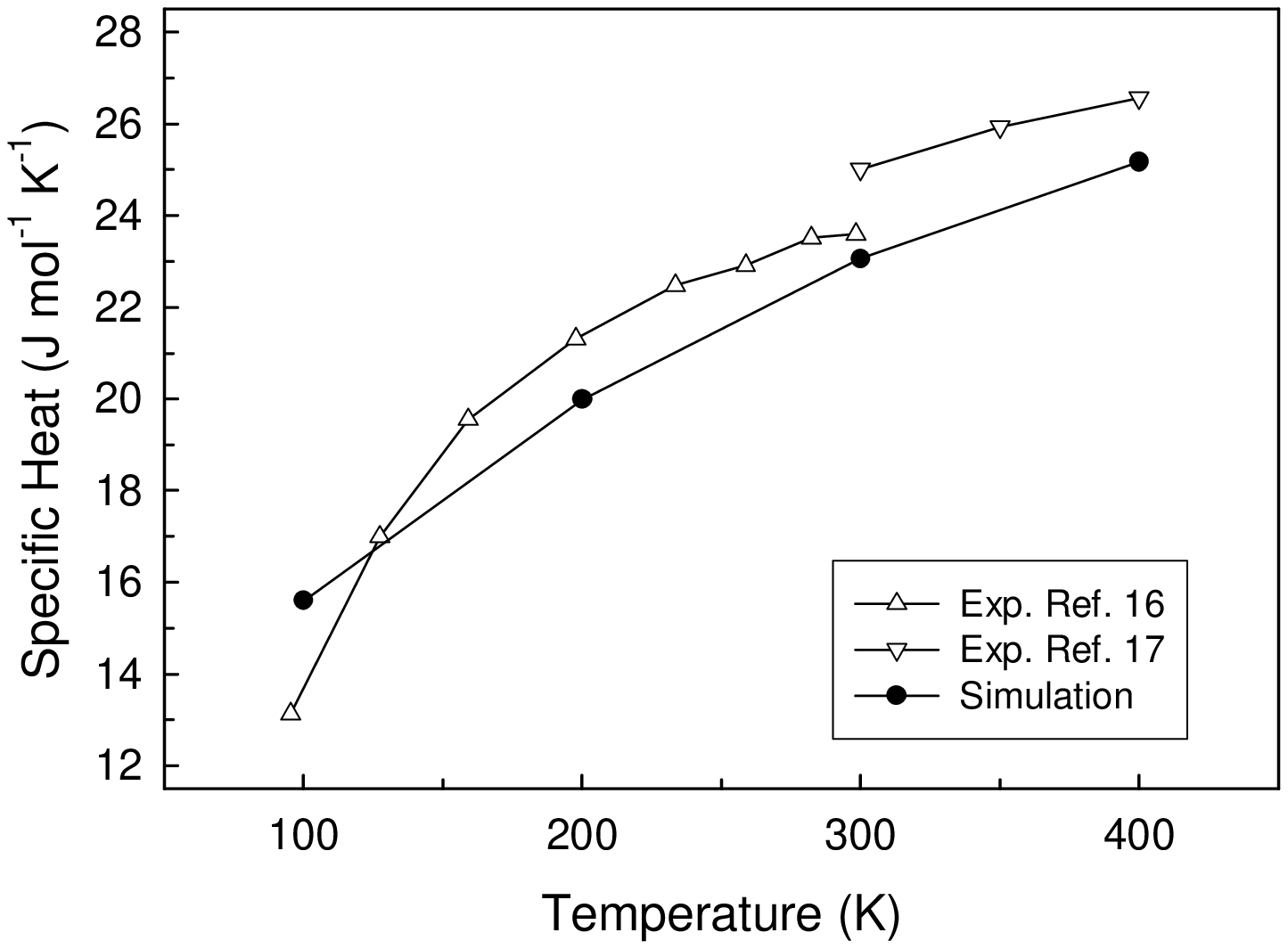}
\caption{Comparison of this simulation and experimental results
for the specific heat capacities of Ti. $\bullet$ present
simulation; $\triangle$, experiment, Ref. \cite{Hun16a};
$\triangledown$, experiment, Ref. \cite{Hun16b}.}
 \label{Hunfig2}
\end{figure}

The temperature development at
different total energies, resulted from the simulation, is shown
in Fig. \ref{Hunfig1}. A linear behaviour and a small increase are seen by
increasing the temperature, although the total energies at each
temperature are constant with small fluctuations. The maximum
fluctuations are roughly $10^{-5}$ eV as expected \cite{Hun2}.
Moreover, the temperature values are observed to be certain within
[0.004, 0.047] K error. The heat capacities are observed by using
the method explained in the preceding section. These are
calculated at each temperature using Eq. \ref{EqHun11}. The results are
compared to the available experimental data \cite{Hun16a,Hun16b} in Fig.
\ref{Hunfig2}. There exists a good agreement. The Maximum difference occurs
at 100 K as of 17\%. The
mean square deviation of the kinetic energy of the atoms in Eq.
(11) is taken as 1s, if it were taken as 2s which covers the 95\%
of the region- they would have certainly covered the whole
experimental values.

\begin{figure}[h] 
\centering
\includegraphics[width=0.8\linewidth,clip]{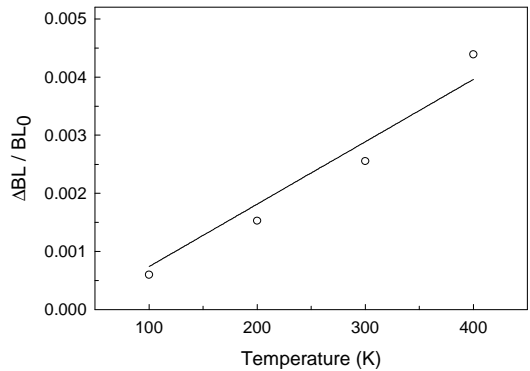}
\caption{The calculated rate of change of the mixed bond length
values against temperature. The solid curve represents the linear
best fit.}
 \label{Hunfig3}
\end{figure}
\subsection{Thermal Expansion Coefficient} The linear
thermal expansion coefficient of the Ti is calculated from Eq.
\ref{EqHun10} using the $MBL$. The $MBL$ varies with the temperature of
the bulk. The temperature dependence of the relative $MBL$ change, $\Delta MBL/MBL_0$, is
shown in Fig. \ref{Hunfig3}, where $MBL_0$ is the mixed bond length at $T=0$ K. It is
important to note that the $T=0$ K value is only a reference point and not a
simulation result. As seen, the curve exhibits a nearly linear
increase in this temperature range.

\begin{table}[h]
\caption{Experimental and our simulation results (in units of
10$^{-6}$/K) for the linear thermal expansion coefficient of
Ti as function of temperature $T$.}\label{Huntab1}
\begin{tabular}{|c|c|c|} \hline
 T (K) &   MD & Experiments \\ \hline \hline 100 & 4.48  &  --- \\ \hline 200 & 7.73
& --- \\ \hline
 & &                        8.35 (a) \\
 & & 8.5   (b) \\
300 & 8.50 & \\
& & 8.6   (c) \\
& & 11.9   (d) \\ \hline 400 &     10.99 & 9.32 (b) \\ \hline
~~300--400~~ &  ~~9.74 (f) ~~ & 8.7   (e) \\ \hline \hline
\multicolumn{3}{|l|}{(a) T= 298.15 K, Ref. \cite{Hun17}} \\
\multicolumn{3}{|l|}{(b) Ref. \cite{Hun11}} \\
\multicolumn{3}{|l|}{ (c) T= 298.15 K, Ref. \cite{Hun12}} \\
\multicolumn{3}{|l|}{(d) T= 23 $^o$C (296.15 K), Ref. \cite{Hun18}} \\
\multicolumn{3}{|l|}{(e) T= 20--100 $^o$C (293.15--393.15 K), Ref.
\cite{Hun17}} \\
\multicolumn{3}{|l|}{(f) Mean value of simulations between $T$ =
300--400 K.} \\  \hline
\end{tabular} \end{table}

The calculated TEC of the Ti values are tabulated in Table
\ref{Huntab1} together with the available experimental results. We
could not find any experimental TEC value below 300 K to compare
with our simulation results. The experimental results at 300 K
\cite{Hun11, Hun12, Hun17, Hun18} show noticeable irregularity.
One of them perfectly matches with the simulation result. At 400
K, the simulation result of $10.99 \times 10^{-6}$/K differs by 18\% from
the experimental result of Ref. \cite{Hun11}. The difference is
smaller in the 300--400 K range, 12\%, where the simulation and
experimental results are $9.74 \times 10^{-6}$/K and $8.7 \times 10^{-6}$/K,
respectively.

\section{Summary} Isobaric and isoenergetic MD simulations have been
performed with Verlet algorithm. The LJ potential parameters of
Yamamoto-Kagawa-Doyama is used for the Ti-Ti interactions. A Ti--389
sample is produced by using the same lattice constants of this
potential. The elastic minimum image convention of the periodic
boundary conditions is set to enable the elasticity of the bulk.
Constant pressure, and isoenergetic simulations have been done,
one after the other. Specific heat and linear TEC of Ti are
calculated between 100 K and 400 K. The lower temperature is
determined to reduce the quantum effects \cite{Hun13}. And the
upper limit is set to the Debye's temperature since the binary
potentials require modifications for the higher temperatures
\cite{Hun3}. The calculated TEC value at 300 K matches perfectly
with one of the experiment, while the highest difference occurred
as 18\% at 400 K. Furthermore, observed path of the specific heat
in the simulation has a similar path with the experiments as shown
in Fig. \ref{Hunfig2}. Numerical convergence problems at both temperature
limits are observed in the simulation. Solving the high
temperature problems will be our future work to clarify the
behaviour in this regime.
\section*{Acknowledgement}
The work was performed at the Institut f\"ur Physik der
Universit\"at Greifswald. Part of this work was supported by the
Istanbul Technical University through Young Researchers
Scholarship, and by the Deutsche Forschungsgemeinschaft (DFG)
through SFB 198.

\end{document}